\documentclass[%
aip,
jmp,%
amsmath,amssymb,
reprint,%
author-year,%
author-numerical,%
]{revtex4-1}

\usepackage{graphicx}
\usepackage{bm}
\usepackage{enumitem} 

\usepackage{caption,subcaption}
\begingroup
\renewcommand{\section}[2]{}

\newcommand\apj{ApJ}
\newcommand\solphys{Solar Phys.}
\newcommand\aap{A\&A}
\newcommand\ssr{Space Sci. Rev.}
\newcommand\apjl{ApJL}

\newcommand{\disp}{\displaystyle}
\newcommand{\mbf}[1]{\mathbf{#1}}
\newcommand{\fpar}[2]{\frac{\partial #1}{\partial #2}}

\begin{document}
\preprint{AIP/123-QED}

\title[]{A viable non-axisymmetric non-force-free field to represent solar active regions}

\author{A. Prasad}
\email{avijeet@prl.res.in }
\author{R. Bhattacharyya}
 \email{ramit@prl.res.in }
\affiliation{Udaipur Solar Observatory, Physical Research Laboratory, Dewali, Bari Road, Udaipur 313 001, India}
\date{\today}

\begin{abstract}
A combination of analytical calculations and vectormagnetogram data are utilized to develop 
a non-axisymmetric  non-force-free magnetic field and asses its viability in describing 
solar active regions. 
For the purpose, we construct a local spherical shell where a planar surface, tangential 
to the inner sphere, represents a Cartesian cutout of an active region. The magnetic field 
defined on the surface is then correlated with magnetograms.
The analysis finds the non-axisymmetric non-force-free magnetic field, obtained by a 
superposition of two linear-force-free fields, correlates reasonably well with magnetograms.
\end{abstract}

\maketitle

An assumption of axisymmetry is almost customary to describe various processes
occurring in the Sun. For instance, many of the solar dynamo models employing spherical polar coordinates
assume axisymmetry \citep{2010LRSP....7....3C}, and so do the models for the large-scale flows 
(differential rotation and meridional flow) on the surface of the Sun \citep{2014SSRv..186..491J}. 
In contrast,  observation of active regions suggest a complete absence of any symmetry in the photospheric field ${\bf{B}}$. 
This is expected, as the convection zone---through which the buoyant magnetic flux tubes rise---being turbulent \citep{2009LRSP....6....4F} 
is devoid of any symmetry. 
The non-linear coupling of 
 ${\bf{B}}$ with other variables in a hydromagnetic description of the solar plasma \citep{2010LRSP....7....3C} guarantees a 
violation of the axisymmetry in all variables, if the magnetic field is non-axisymmetric. 
Nevertheless, magnetic field topologies that are morphologically similar to sunspots 
can be mimicked from a locally axisymmetric field defined in spherical coordinates cf. Figures 4-9 of \citet{1990ApJ...352..343L},
which was further explored in \citet{2014ApJ...786...81P} 
(hereafter PMR14) by using solutions of axisymmetric non-linear-force-free fields to fit the photospheric magnetograms. It is then imperative to 
find non-axisymmetric magnetic field (in spherical geometry) that are morphologically similar to solar active regions, which is the primary objective of 
the paper. 
Secondarily, the calculations utilize a non-force-free 
description of the magnetic field which is congruent to a more realistic representation of the active regions.

Standardly, the magnetic field of a static photosphere is often approximated to be a force-free field where magnetic pressure is balanced by magnetic tension, leading to zero Lorentz force \citep{2012LRSP....9....5W}. 
Strictly, the assumption is more valid at the chromosphere and the lower corona where magnetic pressure dominates over thermodynamic pressure \citep{2001SoPh..203...71G}. There are several 
numerical techniques which extrapolate  three dimensional  force-free magnetic fields from two dimensional photospheric vector magnetograms. Some of the 
contemporary techniques include
Optimization \citep{2004SoPh..219...87W}, Magnetofrictional \citep{1997SoPh..174..191M},  
Grad-Rubin based \citep{ 2011ApJ...728..112W}, and Green's function-based methods \citep{2000SoPh..195...89Y}.
The above techniques have limitations in reproducing the coronal field faithfully because of the following reasons \citep{2012LRSP....9....5W}:
the non force-free nature of the photosphere, unavailability of boundary conditions on all boundaries of a computational box (observational data provides only the bottom boundary),
 uncertainties on vector-field measurements (particularly of the transverse component) and the requirement that a large physical domain needs to be modeled to capture the magnetic connectivity of an active region to its surroundings. The above limitations necessitate an analytical description  where ${\bf{B}}$ is 
non-force-free \citep{2011PhPl...18h4506K, 2007SoPh..240...63B} along with an employment of spherical polar coordinates
 since the required larger physical domain may not necessarily be approximated by a local Cartesian volume \citep{2014A&A...562A.105T}. 
Toward removing some of these limitations, in the paper we analytically explore the relevance of a non-axisymmetric non-force-free magnetic field in describing 
the active regions. Importantly, the  analytical approach  provides explicit non-axisymmetric modes of the magnetic field which is difficult to 
identify from numerical extrapolations.

To keep  
calculations in analytical domain, we skip non-linear-force-free-fields for which only the axisymmetric semi-analytical solutions are available \citep{1990ApJ...352..343L,2014ApJ...786...81P}
and concentrate on the linear-force-free field  $\mbf{B}^{f}$ \citep{1956PNAS...42....1C,1957ApJ...126..457C} satisfying

\begin{equation}
\nabla\times\mbf{B}^{f} = \alpha \mbf{B}^{f}= 0,
\label{e:ff}
\end{equation}
where the constant $\alpha$ represents the magnetic circulation per unit flux \citep{2012PPCF...54l4028P}.
 The linear-force-free field can be interpreted as an eigenvalue equation of the operator curl with solutions forming a complete orthonormal basis \citep{yoshida-giga}. 
The vector ${\bf{B}}^f$ is also nomenclatured as  Chandrasekhar-Kendall (CK) eigenfunction \citep{1957ApJ...126..457C} and in 3D spherical polar coordinates is given by
\begin{equation}
\mbf{B}^f=\frac{1}{\alpha} \nabla\times\nabla\times\psi \mbf{r}+\nabla\times\psi \mbf{r},
\label{e:ck}
\end{equation}
 where ${\bf{r}}$ is the position vector.  The eigenfunction $\psi$ is the solution of the Helmholtz equation $(\nabla^2+\alpha^2)\psi=0$ and is given by
\begin{equation}
\psi_{lm}(r,\theta,\phi)=C_{lm}[C_1 j_l(\alpha r)+C_2y_l(\alpha r)]P_l^m(\cos\theta)\exp(im\phi),
\label{e:psi}
\end{equation}

\noindent  where $j_l(r,\theta,\phi)$ and $y_l(r, \theta,\phi)$ represent the spherical Bessel functions of the first and second kind respectively.
The components of the magnetic field are given by
\begin{subequations}
\begin{align}
B^f_r(r,\theta,\phi)&=\frac{-1}{\alpha r}\left[\frac{1}{\sin\theta}\fpar{}{\theta}\left(\sin \theta \fpar{\psi}{\theta}\right)+\frac{1}{\sin^2\theta}\fpar{^2\psi}{\phi^2}\right], \label{br}\\
B^f_\theta(r,\theta,\phi)&=\frac{1}{\alpha r}\fpar{}{r}\left(r\fpar{\psi}{\theta}\right)+\frac{1}{\sin\theta}\fpar{\psi}{\phi},\label{bt}\\
B^f_\phi(r,\theta,\phi)&=\frac{1}{\alpha r \sin \theta}\fpar{}{r}\left(r\fpar{\psi}{\phi}\right)-\fpar{\psi}{\theta}\label{bp}.
\end{align}
\label{bf}
\end{subequations}
 Noteworthy is the scale-independence of $|\mathbf{B}^f|$.

\begin{figure}[h!]
  \centering
    \includegraphics[width=0.85\linewidth]{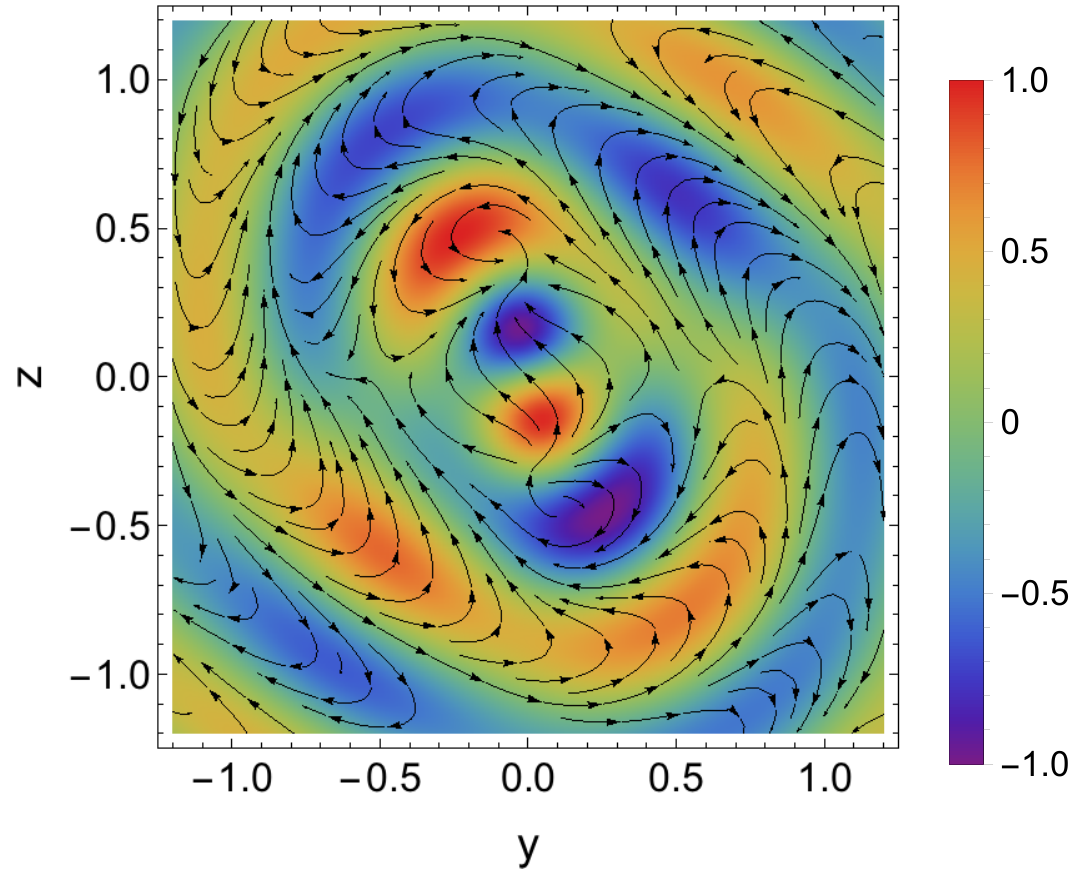}
    \caption{Cross-section of non-axisymmetric linear-force-free field corresponding to $l=3$, $m=2$ and $\alpha =9$ at $x=0.5$ . The density plot represents the strength of the vertical magnetic field while the surface components of the magnetic field are represented by the stream lines, with arrows depicting the direction of the field.}
  \label{f:merck}
\end{figure}

Toward developing the non-axisymmetric non-force-free field, notable are the following points:

\begin{enumerate}[label=(\roman*)]
\item  A general non-force-free field, relevant to the solar corona, can be obtained by superposing two linear-force-free fields 
 having two different eigenvalues \citep{2011PhPl...18h4506K, 2007SoPh..240...63B}. 

\item  To correlate the above non-force-free field with magnetograms, initially the linear-force-free fields are to be calculated in a spherical shell of 
inner radius ($r_0$) and outer radius ($r_1$), where a planar surface tangent to the boundary at radius $r_0$ represents a part of the 
photosphere. Further the outer boundary $r_1$ has to be 
a magnetic flux surface to avoid the nonphysical scenario where magnetic field lines (MFLs) extend to infinity and the asymptotic magnetic energy becomes infinite \citep{1978SoPh...58..215S}.
The condition that the field lines are enclosed within the shell, using equation \eqref{e:psi}, requires $C_1 j_l(\xi_1)+C_2 y_l(\xi_1)=0$ where $\xi_1=\alpha r_1$;
which gives $\disp{\frac{C_1}{C_2}=-\frac{y_l(\xi_1)}{j_l(\xi_1)}}$. Absorbing the constants in $C_{lm}$ of equation \eqref{e:psi}, such that $\disp{A_{lm}=\frac{-C_{lm} C_2}{j_n(\xi_1)}}$, we arrive at 
the following expression for the $\psi$ 
\begin{align}
\psi_{lm}(\xi,\theta,\phi)=&A_{lm}[ j_l(\xi)y_l(\xi_1)-j_l(\xi_1)y_l(\xi)]\nonumber\\
&\times P_{lm}(\cos\theta)\exp(im\phi).
\end{align}
The corresponding magnetic field can be calculated utilizing equations (\ref{bf}). The axisymmetric mode is characterized by $m = 0$ whereas $m>0$ gives the 
non-axisymmetric modes.
A meridional cross-section of the linear-force-free field for $l=3$, $m=2$ and $\alpha =9$ is shown in Figure \ref{f:merck}, where the non-axisymmetric nature is markedly visible.

\end{enumerate}

\noindent The non-force-free field within the spherical shell is then given by 
\begin{equation}
\mathbf{B'} = \mathbf{B}^f_1 + \mathbf{B}^f_2;~~~ {\rm{with ~Lorentz~ force}} \quad \mathbf{J'}\times \mathbf{B'}=\epsilon \alpha_1 \mathbf{B}^f_2\times \mathbf{B}^f_1,
\label{e:nff}
\end{equation} 

\noindent where $\mathbf{B}^f_1$ and $\mathbf{B}^f_2$ are the two linear-force-free fields defined by equations (\ref{bf}). 
The corresponding  eigenvalues are written as $\alpha_1$ and $\alpha_2 =(1+\epsilon)\alpha_1$ respectively such that  the 
Lorentz force (in usual notations) $\mathbf{J'}\times\mathbf{B'}=0$  for $\epsilon=0$. The $\epsilon$ then quantifies the 
non-force-free field $\mathbf{B'}$. For finding
optimal non-force-free states, we restrict $\epsilon$ in the range $0<\epsilon<1$.

To asses the applicability of the aforementioned non-force-free field $\mathbf{B'}$ to active regions, we select the procedure developed in 
\citet{2014ApJ...786...81P}. The selection is based on the advantages gained in terms of 
obtaining fast and reasonably good fits to observed vector magnetograms while restricting the calculations in analytical domain. 
With the details in PMR14, here we mention the 
salient features of the procedure.  The procedure employs generation of 2D template vector magnetograms from the analytical 3D solutions presented in equations \eqref{bf} - \eqref{e:nff},  which 
are then fitted to the observed vector magnetograms. We then take a cross section of a sphere at a inner radius $r_0$, and compute all three components of magnetic field  over this 
2D surface. The orientation of the magnetogram is varied through two Euler rotations $\theta'$ and $\psi'$; cf. Figure 4 of PMR14 for details. 
Thus, the parameter space to look for a best fit ($\mathbf{B}_T$; representing the theoretical field under consideration) with the magnetic field ($\mathbf{B}_O$) from magnetogram comprises 
of different modes $l$, $m$ and variables $\alpha$, $r_0$, $r_1$, $\theta'$, $\psi'$ which are
obtained by maximizing the correlation parameter $c$, where

\begin{equation}
  c=\frac{\langle(\mathbf{B}_T\cdot\mathbf{B}_O)| \mathbf{B}_O|\rangle}{\langle|\mathbf{B}_T|^3
\rangle^{1/3}\langle|\mathbf{B}_O|^3\rangle^{2/3}},
\label{fitting}
\end{equation}
\noindent represents the grid-averaged normalized scalar product between the two vectors weighted by the strength
of the observed magnetic field. The value of $c$ lies between 0 and 1 with 1 representing a perfect correlation.

\begin{figure}[hp]
  \centering
  \begin{subfigure}[]{0.36\textwidth}
    \centering
    \includegraphics[width=1\linewidth]{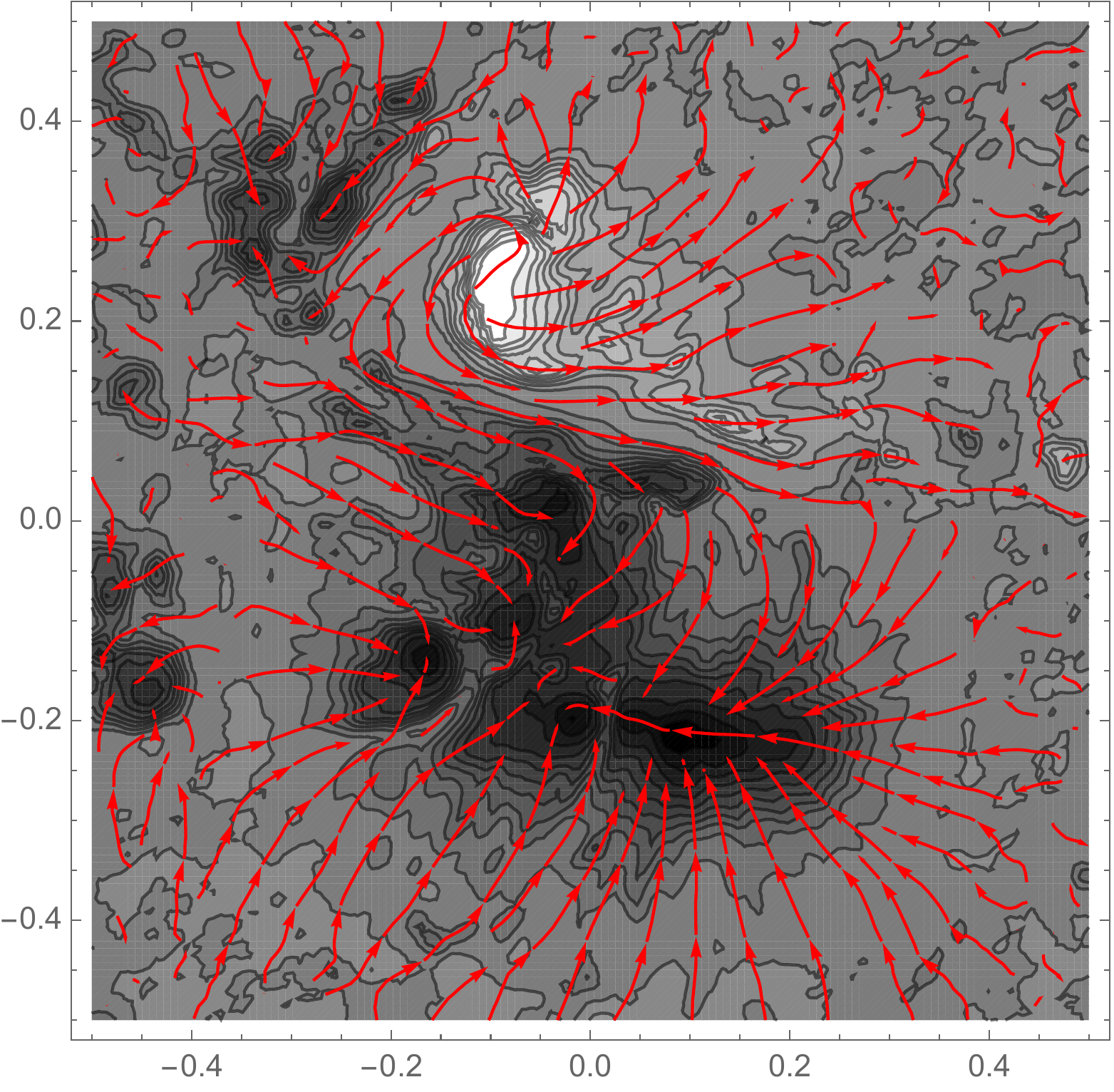}
    \caption{}
    \label{f:obs}
  \end{subfigure}
  \quad
  \begin{subfigure}[]{0.36\textwidth}
    \centering
    \includegraphics[width=1\linewidth]{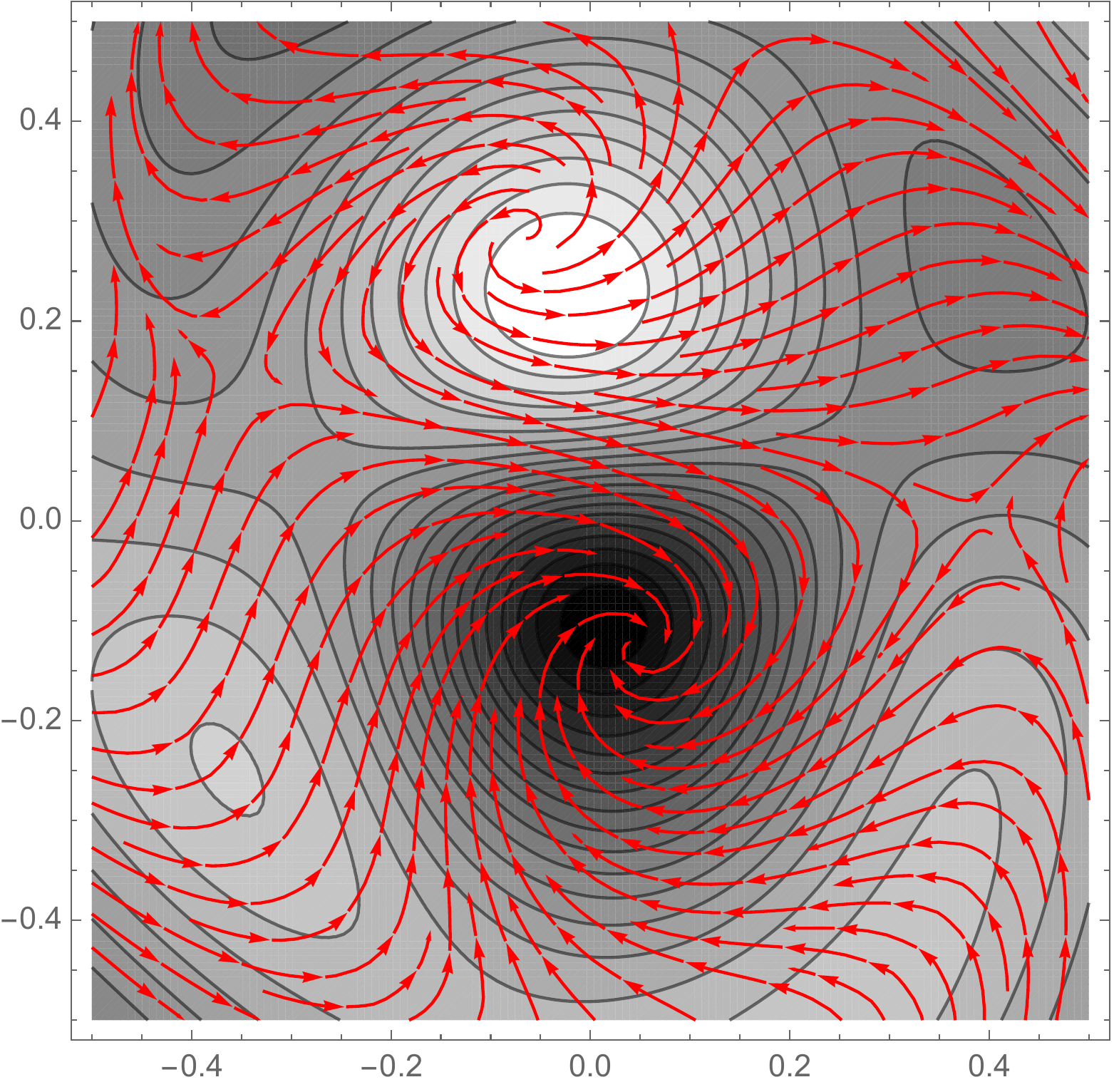}
    \caption{}
    \label{f:nff}
  \end{subfigure}
\begin{subfigure}[]{0.36\textwidth}
    \centering
    \includegraphics[width=1\linewidth]{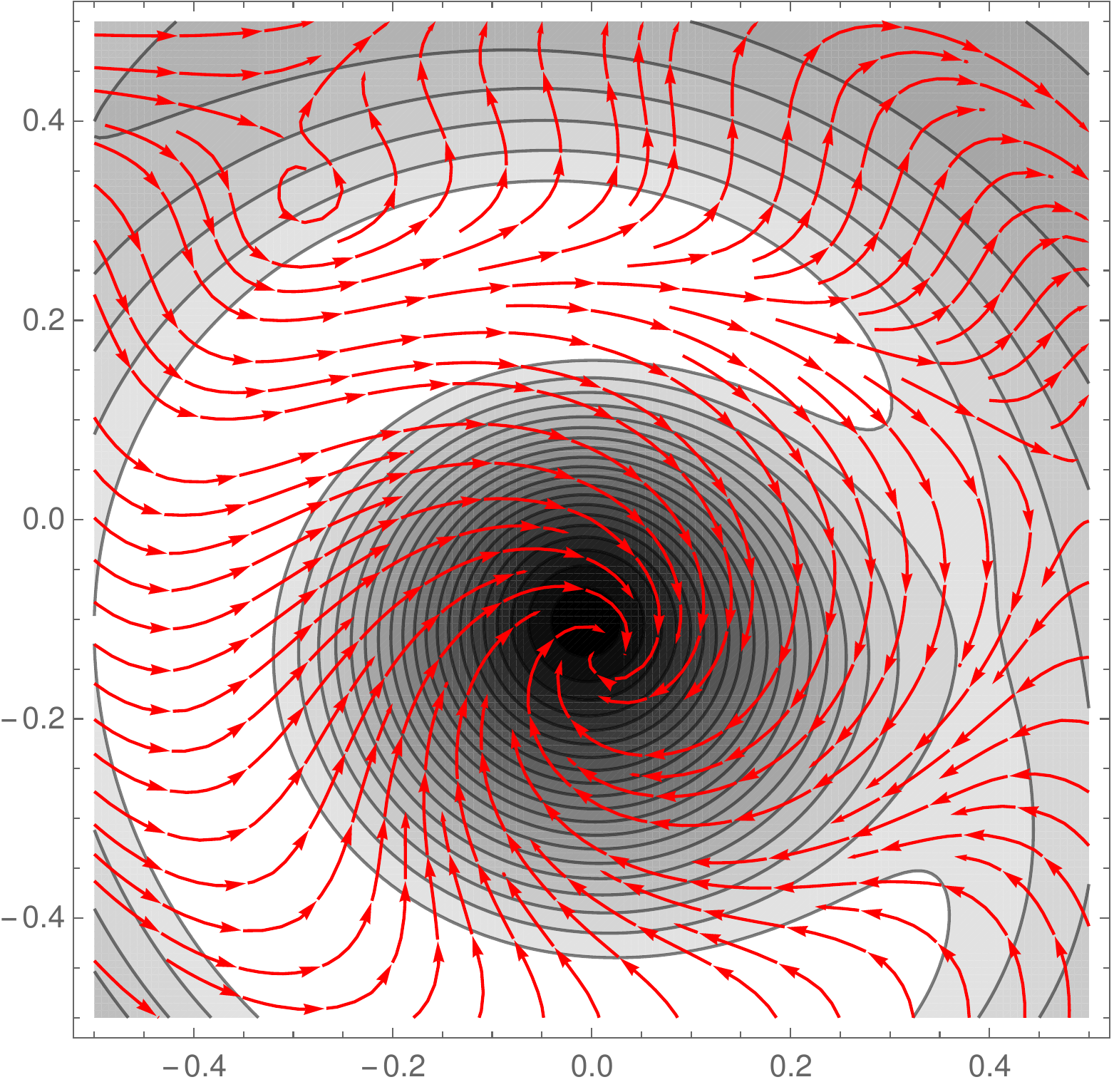}
    \caption{}
    \label{f:nff}
  \end{subfigure}
  \caption{ Vector magnetogram for AR 11283 on September 7, 2011 at 02:00 hours represented using (a) observational data from HMI/SDO and (b) the  best-fit non-axisymmetric non-force-free case
 (c) the best-fit axisymmetric non-force-free case. The parameters for the best-fit are given in Table 1. The plot description is similar to that of Figure 1. }
  \label{f:fits}
\end{figure}

We choose the vector magnetogram of active region (AR) NOAA 11283 observed on September 7, 2011 at 02:00 hours from the Heliospheric Magnetic Imager (HMI) \citep{2012SoPh..275..229S} on board the Solar Dynamics 
Observatory (SDO) \citep{2012SoPh..275....3P}. The full-disk vector magnetograms from HMI have a spatial resolution of $0''.5$ per pixel and a temporal cadence of 12 minutes. HMI samples the Fe I 6173 $\AA$ spectral line at six 
different wavelengths for six polarization states (I $\pm$ S, where S = Q, U, and V). The Stokes parameters, I, Q, U, and V are inverted through the Very Fast Inversion of the Stokes Vector code (VFISV) \citep{2011SoPh..273..267B} 
which is based on the Milne-Eddington atmosphere. The 180$^\circ$ambiguity is resolved by using the minimum energy method \citep{1994SoPh..155..235M,2009SoPh..260...83L}.  AR 11283 produced several 
energetic flares and CMEs over the period of a week after appearing on September 1, 2011 \citep{2014ApJ...795..128L}. The magnetic topology of the AR is complex in terms of having 
strong bipolar MFLs and diffused regions of weaker fields (Figure \ref{f:obs}).

\begin{table}[]
\centering
\resizebox{0.5\textwidth}{!}{%
\begin{tabular}{|c|c|c|c|c|c|c|c|c|c|}
\hline
Sl no. & model                 & correlation & $\alpha_1,\alpha_2$ & $l$ & $m$ & $r_0$       & $r_1$ & $\theta'$ & $\psi'$                            \\ \hline
1      & Non-axisymmetric non-force-free field    & 0.62    & 9, 10.35    & 3   & 2   & 0.37 & 1     & 1.77 & 0.2                          \\ \hline
2      & Axisymmetric non-force-free field        & 0.60    & 9, 10.35    & 3   & 0   & 0.40 & 1     & 2.75 & 0.4                          \\ \hline
3      & Non-axisymmetric linear-force-free field & 0.62    & 9, 0        & 3   & 2   & 0.4  & 1     & 1.77  & $0 $\\ \hline
4      & Axisymmetric linear-force-free field     & 0.50    & 9, 0        & 3   & 0   & 0.4  & 1     & 0.79 & $0$ \\ \hline
\end{tabular}%
}
\caption{Correlations for non-axisymmetric/axisymmetric non-force-free and non-axisymmetric/axisymmetric linear-force-free fields. The corresponding parameters $\alpha_1$, 
$\alpha_2$, $l$ and $m$ fix a given mode while $r_0$, $r_1$, $\theta'$ and $\psi'$ determine the computation domain.}
\label{t:cor}
\end{table}

\begin{figure}[h!]
  \centering
  \begin{subfigure}[]{0.45\textwidth}
    \centering
    \includegraphics[width=1\linewidth]{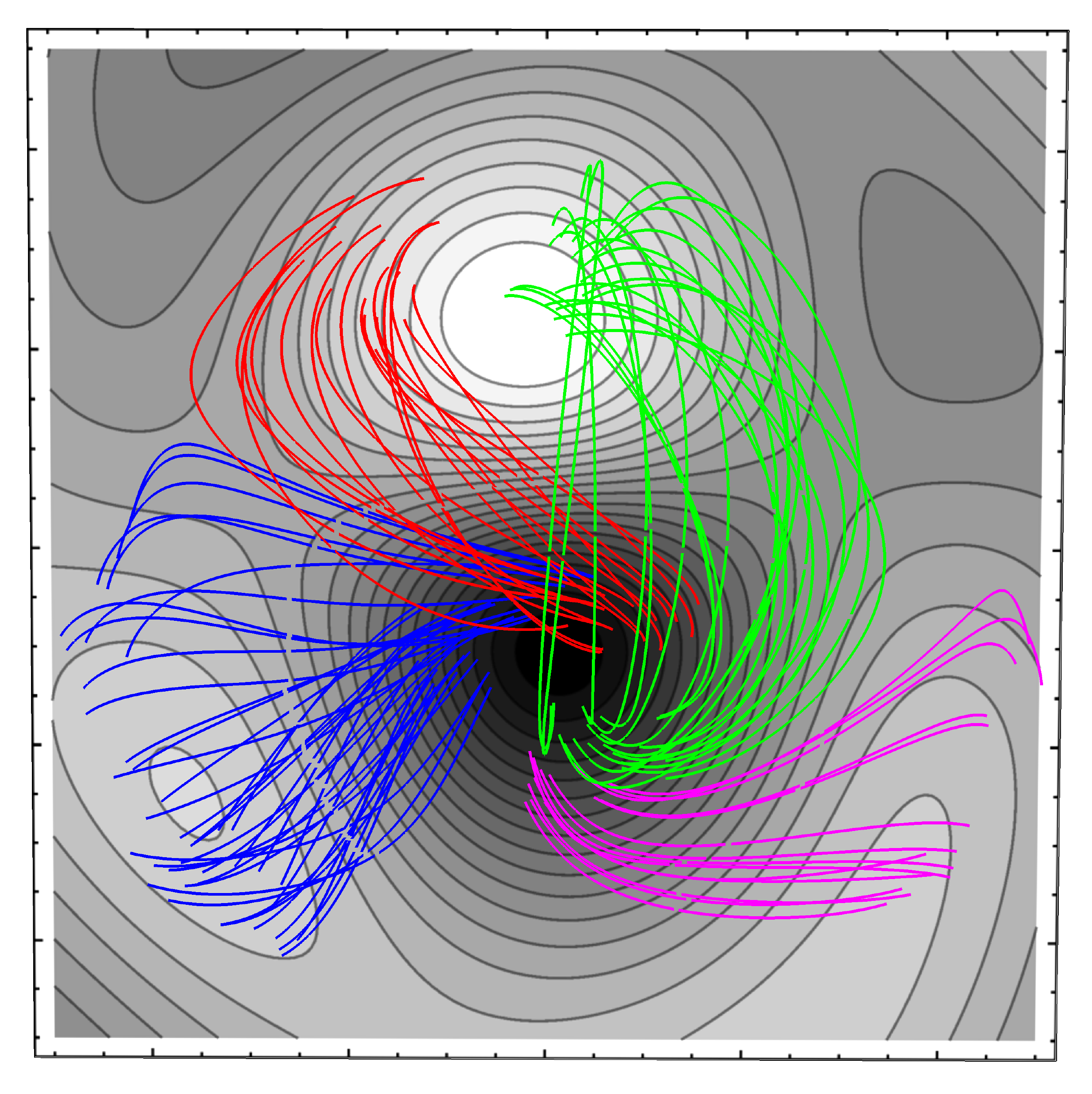}
    \caption{}
    \label{f:lines}
  \end{subfigure}
  \quad
  \begin{subfigure}[]{0.5\textwidth}
    \centering
    \includegraphics[width=1\linewidth]{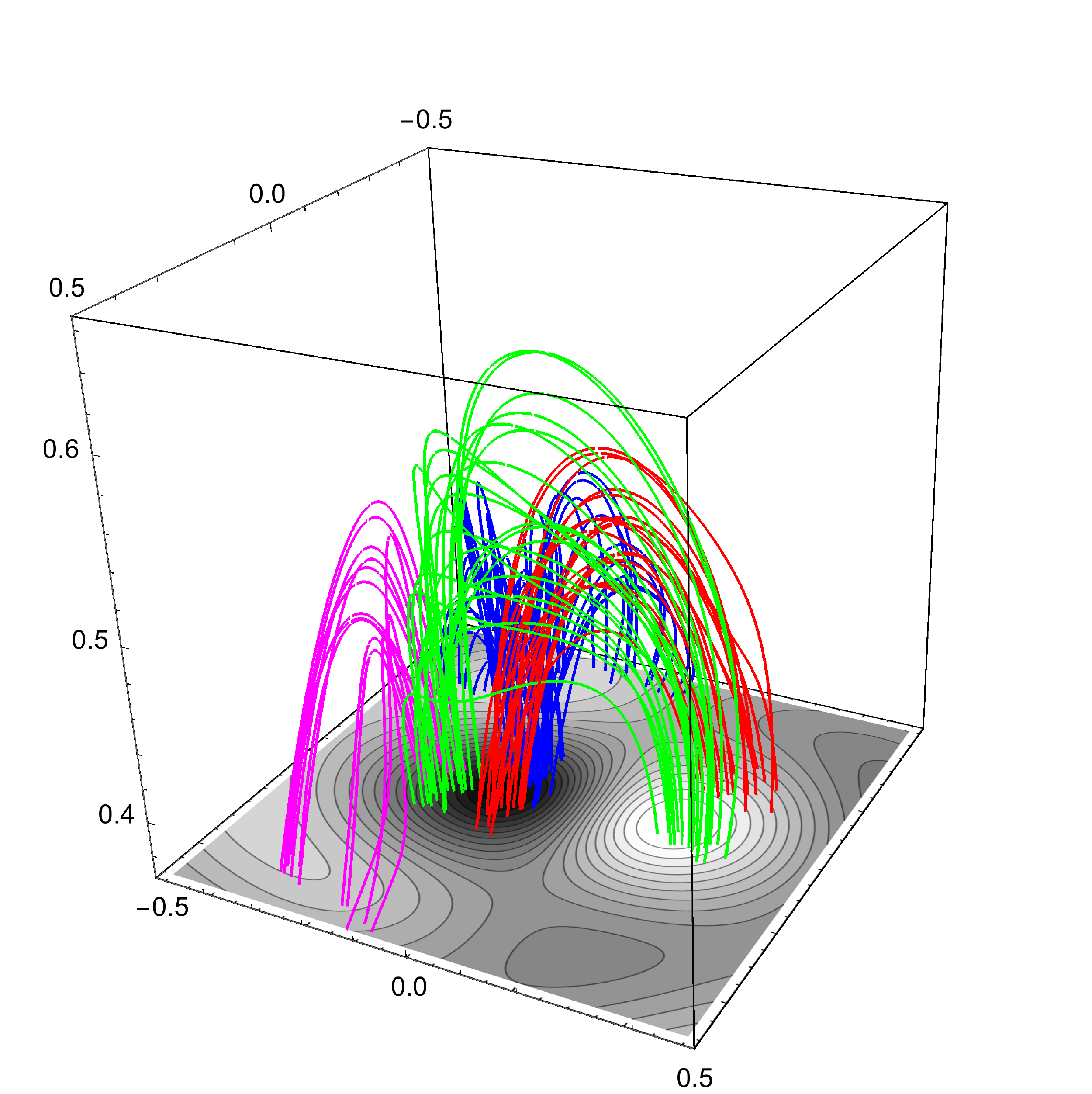}
    \caption{}
    \label{f:lines3d}
  \end{subfigure}
  \caption{ Magnetic field lines of the best fitted non-axisymmetric non-force-free field (cf. Table 1). Panels  (a) and (b) depict the top and the side views respectively. Important is the existence of possible QSLs located in the void between the red and blue, and, the green and
magenta colored field lines.}
  \label{f:global}
\end{figure}

\noindent In Table 1  we list the best correlation of the
 non-axisymmetric and axisymmetric non-force-free field with the magnetogram along with the corresponding parameter set. 
As a reference, we also provide the same for the non-axisymmetric/axisymmetric linear-force-free fields. The maximum correlation is always larger in a
non-axisymmetric mode compared to the axisymmetric mode of a given field.
Importantly, the correlation for the non-axisymmetric non-force-free field is reasonably
good and the magnetic field (shown for grid resolution of 140 $\times$ 140 pixels; cf. Figure  \ref{f:fits})  
is morphologically similar to the vector magnetogram whereas the corresponding axisymmetric mode shows no such similarity. 
Notably, the maximum correlation is identical for the non-force-free and the linear-force-free fields but the photosphere being non-force-free,
within the used analytical framework the non-axisymmetric non-force-free field is more appropriate to represent the vector magnetogram. Moreover, the 
MFLs for the non-force-free field (Figure \ref{f:global}) indicate a possible existence of two quasi-separatrix layers (QSLs) \citep{1996A&A...308..643D} located at the void between the red and blue 
colored field lines, and, the green and magenta colored field lines. Importantly, the group of field lines situated in close proximity on either side of the voids connect to two entirely different 
locations on the grid and results in development of current-sheets \citep{2015PhPl...22h2903K}. The consequent magnetic reconnections may trigger onset of flares. Interestingly, for the selected AR,  
Extreme Ultra Violet brightening  occurs at the general neighborhood of the QSLs 
before onset of the two X-class flares on September 6 and 7, 2011 \citep{2013ApJ...771L..30J}.

In retrospect, the importance of the work is in finding analytical non-axisymmetric magnetic fields which correlates well with magnetograms. The finding, obviously, depends on the  magnetic field 
used to model the photosphere and assumptions inherent to analytical methods. Additionally,  the non-force-free magnetic field generated by superposing two linear-force-free fields is found to fit 
reasonably well with the observed data and  morphologically resembles the 
photospheric magnetic field.   The existence of QSLs is also suggestive of magnetic reconnections which is further supported by the two 
X-class flares  occurring in their general vicinity and warrants further research.

 Data and images are courtesy of NASA/SDO
and the HMI and AIA science teams. SDO/HMI is a joint effort of
many teams and individuals to whom we are greatly indebted for
providing the data. The authors thank an anonymous referee for 
 constructively  criticizing the paper.

\nocite{*}
\bibliographystyle{aipauth4-1}
\bibliography{ms}

\end{document}